\documentclass{svjour2}                    
\smartqed  
\usepackage{graphicx}
\usepackage{amsfonts}
\usepackage{amssymb}
\usepackage{amsmath}
\usepackage{float}
\usepackage{epstopdf}
\begin{document}

\title{Exotic Ground States of Directional Pair Potentials via Collective-Density Variables
}


\author{Stephen Martis         \and
        \'Etienne Marcotte \and
         Frank H. Stillinger \and
         Salvatore Torquato 
}


\institute{S. Martis \at
              Department of Physics, Princeton University, 
Princeton New Jersey 08544, USA \\
           \and
           \'E. Marcotte \at
           Department of Physics, Princeton University, Princeton New Jersey 08544, USA\\
           \and
           F.H. Stillinger \at
           Department of Chemistry, Princeton University, Princeton, New Jersey 08544, USA\\
           \and
           S. Torquato \at
              Department of Physics, Princeton University, Princeton, New Jersey 08544, USA\\
              Department of Chemistry, Princeton University, Princeton, New Jersey 08544, USA\\
              Princeton Center for Theoretical Science, Princeton University, Princeton, New Jersey 08544, USA\\
              Princeton Institute for the Science and Technology of Materials, Princeton University, Princeton, New Jersey 08544, USA\\
              Program in Applied and Computational Mathematics, Princeton University, Princeton, New Jersey 08544, USA\\
              \email{torquato@electron.princeton.edu}
}

\date{Received: date / Accepted: date}

\maketitle

\begin{abstract}
Collective-density variables have proved to be a useful tool in the prediction and manipulation of how spatial patterns form in the classical many-body problem.  Previous work has employed properties of collective-density variables along with a robust numerical optimization technique to find the classical ground states of many-particle systems subject to radial pair potentials in one, two and three dimensions.  That work led to the identification of ordered and disordered classical ground states.  In this paper, we extend these collective-coordinate
studies by investigating the ground states of directional pair potentials in two dimensions.  Our study focuses on directional potentials whose Fourier representations are non-zero on compact sets that are symmetric with respect to the origin and zero everywhere else.  We choose to focus on one representative set which has exotic ground-state properties: two circles whose centers are separated by some fixed distance.  We obtain ground states for this ``two-circle" potential that display large void regions in the disordered regime.  As more degrees of freedom are constrained the ground states exhibit a collapse of dimensionality characterized by the emergence of filamentary structures and linear chains. This collapse of dimensionality has not been observed before in related studies.
\end{abstract}

\section{Introduction}
\label{intro}

The Fourier components of a set of position or density variables are commonly referred to as collective-density variables and the resultant space is called the collective-coordinate space.  The sort of analysis involving collective-density variables naturally arises in the context of the $N$-body problem, where a collective-coordinate transformation will often simplify seemingly intractable problems.  The reader is referred to Feynman's discussion of the superfluid properties of $^4$He or a series of papers of Percus and Yevick for illustrative examples of their use \cite{Feynman72,Yevick56,Percus58}.

Recently, there has been renewed interest in collective-density variables and the
associated ground states for a class of soft, long-ranged
pair potentials  \cite{Fan91,Uche04,Suto05,Suto06,Uche06,Torquato08,Batten08,Torquato11,Zachary11}.  
A proof of the existence of periodic ground states for this class of soft, long-ranged, oscillatory pair potentials utilized such an approach \cite{Suto05,Suto06}.  Duality relations were derived, connecting these ground states to those of particle 
systems subject to certain short-ranged, soft potentials, allowing a rigorous understanding
of the ground states of such many-particle systems \cite{Torquato08,Torquato11}. Various studies 
utilized properties of collective-density variables to demonstrate the existence of both disordered and crystalline 
ground-state configurations for classical many-particle systems subject to isotropic pair potentials 
in one, two and three dimensions \cite{Fan91,Uche04,Uche06,Batten08,Zachary11}.   These interactions also exhibit unusual 
excited states; see Ref. \cite{Batten11} and references therein.

The present paper extends these last collective-coordinate investigations to such classical systems in $\mathbb{R}^2$ 
subject to anisotropic or directional potentials.  As shall be seen, the introduction of anisotropy results in several striking, unintuitive features of the ground state configurations.  In the disordered regime, ground states display a tunable void structure.  Ground states exhibit a collapse of dimensionality characterized by the emergence of filamentary structures as more collective density variables are minimized.

It is useful to remark on the possible applications of the results presented here.  The pair potentials we study are directional in addition to being long-ranged, oscillatory, and bounded at the origin, characteristics which are in some sense artificial.   However, such interactions might present themselves in polymer systems \cite{Likos06,Zachary08,Flory50} or in systems of particles adsorbed on a metallic substrate in which the conduction electrons directionally interact with the adsorbed material \cite{Venables00}.  Furthermore,  
a decoration of {\it disordered} classical ground-state configurations generated from
the collective-coordinate approach has  found real-world applications in the 
design of disordered dielectric metamaterials with large and complete photonic band gaps \cite{Florescu09}.

In the present paper we use "directional interaction" to refer to pair potentials between point particles that depend on direction relative to
a fixed spatial coordinate system.  Thus, we do not consider cases of anisotropic particles with rotational degrees of freedom, such as ``patchy
colloids"  \cite{Giacometti10,Bianchi11}.


\section{Definitions and Preliminaries}
\label{def}

We proceed to define the collective-density variables of a particle distribution and elucidate some of their important properties.  We are given a configuration of $N$ point particles confined to a volume $\Omega$ in $d$-dimensional Euclidean space $\mathbb{R}^d$, subject to periodic boundary conditions.  The local particle density of the system can be expressed as a sum of delta functions:

\begin{equation}
\rho(\mathbf{r}^N) = \sum_{j=1}^{N}\delta(\mathbf{r}-\mathbf{r}_j)\textrm{.}
\end{equation}
The collective-density variables are then conventionally defined as:
\begin{equation}
\tilde{\rho}(\mathbf{k}) = \sum_{j=1}^{N}\textrm{exp}(i\mathbf{k\cdot r}_j)\textrm{.}
\end{equation}
which, to within a constant factor, are the Fourier components of Eq. (1), where $\mathbf{r}_j$ denotes the $j^\textrm{th}$ particle position and $\mathbf{k}$ is a wave vector defined by the parameters of the $d$-dimensional fundamental cell.  For a rectangular cell:
\begin{equation}
\mathbf{k} = \bigg(\frac{2\pi n_1}{L_1}, \frac{2\pi n_2}{L_2},...\,,\frac{2\pi n_d}{L_d}\bigg),\;\;\;\; n_1,...\,,n_d\in\mathbb{Z}.
\end{equation}  
In principle, there are an infinite number of $\tilde{\rho}(\mathbf{k})$ at our disposal.  However, since the system only possesses $dN$ configurational degrees of freedom, the point pattern is completely described by $dN$ collective-density variables (although it has been shown that as few as half as many may completely constrain a system, at least in one dimension \cite{Fan91}).  

There are some properties of the $\tilde{\rho}(\mathbf{k})$, stated now, which will be useful in the following analysis.  For the zero $\mathbf{k}$-vector, the corresponding collective-density variable takes on a value independent of particle positions, which is simple to compute from the definition above:
\begin{equation}
\tilde{\rho}(\mathbf{0}) = N.
\end{equation}
For all other $\tilde{\rho}(\mathbf{k})$, we have the following properties which follow from the relations described above:
\begin{eqnarray}
\tilde{\rho}(\mathbf{-k}) =\tilde{\rho}^*(\mathbf{k})\\
0\leq|\tilde{\rho}(\mathbf{k})|\leq N.
\end{eqnarray}
As will be made clear in the following development, it will be useful to define the ``real collective-density variables" in the following way:
\begin{eqnarray}
C(\mathbf{k})&=&\frac{|\tilde{\rho}(\mathbf{k})|^2-N}{2}\\
		     &=&\sum_{j<l=1}^N\textrm{cos}[\mathbf{k}\cdot(\mathbf{r}_j-\mathbf{r}_l)].
\end{eqnarray}
The real collective-density variables have the following properties:
\begin{eqnarray}
C(\mathbf{0})&=&\frac{N(N-1)}{2}\\
C(-\mathbf{k})&=&C(\mathbf{k})\\
-\frac{N}{2}\leq &C(\mathbf{k})&\leq\frac{N(N-1)}{2}.
\end{eqnarray}

We can also relate the real collective-density variables to the structure factor $S(\bf{k})$ which describes how a particle configuration scatters incident radiation.  In terms of the real collective-density variables, $S(\mathbf{k})$ is shown to be \cite{Uche04}:
\begin{equation}
S(\mathbf{k}) = \frac{|\tilde{\rho}(\mathbf{k})|^2}{N} = 1+\frac{2C(\mathbf{k})}{N}.
\end{equation}

We now introduce $v(\mathbf{r})$, a pair potential which is symmetric under inversion and translationally invariant.  In addition, we assume that its Fourier transform, $V(\mathbf{k})$ is well-defined, such that:
\begin{eqnarray}
V(\mathbf{k}) &=& \int_\Omega d\mathbf{r}\;\;v(\mathbf{r})e^{i\mathbf{k\cdot r}}\\
v(\mathbf{r}) &=& \frac{1}{\Omega}\sum_\mathbf{k}V(\mathbf{k})e^{-i\mathbf{k\cdot r}}.
\end{eqnarray}
Then, the total interaction energy of the system is given by the sum of the interaction potential over all particle pairs, which we can relate to the $C(\mathbf{k})$:
\begin{eqnarray}
\Phi(\mathbf{r}^N) &=& \sum_{j<l=1}^N v(\mathbf{r}_j-\mathbf{r}_l)\\
			     &=& \frac{1}{2\Omega}\sum_\mathbf{k}V(\mathbf{k})[\tilde{\rho}(\mathbf{k})\tilde{\rho}(\mathbf{-k})-N]\\
			     &=& \frac{1}{\Omega}\sum_\mathbf{k}V(\mathbf{k})C(\mathbf{k}).
\end{eqnarray}

We proceed to choose an interaction whose Fourier transform is strictly positive on a finite set of wave vectors, $A$, and zero everywhere else.  We then observe that by minimizing the $C(\mathbf{k})$ for all $\mathbf{k}\in A$, we will obtain an absolute minimum for the total energy of the system.  As shown earlier, the $C(\mathbf{k})$ are inversion symmetric, so constraining $C(\mathbf{k})$ to some value necessarily constrains $C(-\mathbf{k})$ to that same value.  In constructing our optimization procedure, then, it is only necessary to consider half of $\mathbf{k}$-space and to define our potential on one half of $\mathbf{k}$-space.  We can then relate the number of constrained $C(\mathbf{k})$ to the total degrees of freedom of the system, defining a dimensionless quantity \cite{Uche04}:
\begin{align*}
\chi = \frac{M}{dN},
\end{align*}
where M is the total number of \emph{independent} $C(\mathbf{k})$ constrained, $d$ is the dimensionality of the particle container, and $N$ is the number of particles.

\section{Computational methods}

In all of the following simulations, we examine particle configurations in $\mathbb{R}^2$ confined to square fundamental cells, the dimensions of which have been normalized so that the total particle density is unity.  Periodic boundary conditions are imposed on the system.  In order to remain consistent with previous work, we try to limit our simulations to $N=418$ particles, when appropriate.  This number was chosen for preliminary studies since 418 particles can be arranged into a nearly undeformed triangular lattice when confined to a square fundamental cell \cite{Uche04}.

In most cases, the initial conditions for the simulations are Poisson distributed configurations drawn from the same distribution.  For these instances, the simulation procedure is analogous to an ``infinite quench" in which the system is brought from an essentially infinite temperature (and a maximally random configuration) to zero temperature (and an energy-minimizing configuration), instantaneously.  For $\chi>0.5$, convergence to an absolute minimum from Poisson initial conditions becomes much more difficult (with only about one in twenty runs converging in the best cases).  In order to ensure convergence to a global minimum, randomly perturbed lattice patterns chosen by ansatz were used as initial conditions.

The minimization itself was accomplished with an implementation of the MINOP algorithm \cite{Kaufman99}.  It was shown in previous work that MINOP is advantageous when compared with other optimization schemes \cite{Batten08}.  The conjugate gradient method was used in preliminary trials, but was later shown to be much less likely to find a global minimum, although the procedure would terminate in marginally fewer steps using this method \cite{Uche06}.

Because of errors resulting from the numerical imprecision of the computer simulations, ground-state energies were never found to be exactly minimal in practice.  The cutoff for determining whether or not a configuration was a ground state or not was set to be $10^{-10}$ natural units from the minimum [which can be calculated from Eqs. (11) and (17) and the form of the potential], although energies were predominantly much closer to the minimum than this (on the order of $10^{-20}$ natural units).  

\section{Directional Pair Potentials}

Previous work has characterized several results for isotropic potentials in two and three dimensions \cite{Uche04,Uche06,Batten08}, which correspond to finite, hyperspherical regions in $\mathbf{k}$-space on which $V(\mathbf{k})$ is a positive constant (specifically set to unity in our simulations in order to simplify analysis), i.e.:
\begin{equation}
V(\mathbf{k}) = \Bigg\{\begin{matrix}
&1&,&\textrm{   }0\leq&|\mathbf{k}|&\leq K,\\
&0&,&\textrm{   }K<&\,\,|\mathbf{k}|,&
 \end{matrix}
\end{equation}
where $K$ is some positive constant. Any ground-state configuration
for this isotropic pair interaction has been termed ``stealthy" \cite{Batten08} because
the associated structure factor $S(k)$ is exactly zero for $k \le K$, i.e.,
single scattering of incident radiation is completely suppressed for $k \le K$.
Moreover, by construction, such stealthy configurations are also ``hyperuniform" \cite{To03},
since infinite-wavelength density fluctuations vanish. In what follows, we will refer to this region on which $V(\mathbf{k})$ is non-zero as the ``$\mathbf{k}$-space exclusion region," since $C(\mathbf{k})$ (or, equivalently, $S(\mathbf{k})$) is minimized on these wave vectors.  In the infinite-volume limit, as $\mathbf{k}$-space becomes dense, the real space potential becomes:
\begin{equation}
v(r) = \frac{V_oK}{2\pi r}J_1(Kr),
\end{equation}
where $J_1$ is the first order Bessel function of the first kind and $r$ denotes the radial distance from the particle.  This potential is both finite at $r=0$ and long-ranged and oscillatory and as $r\rightarrow\infty$:
\begin{equation}
v(r\rightarrow\infty)\sim\frac{V_oK^{1/2}}{\sqrt{2\pi^3r^3}}\textrm{cos}\bigg(Kr-\frac{3\pi}{4}\bigg).
\end{equation}

Utilizing this class of potentials, Uche \emph{et al.} \cite{Uche04} categorized the emergence of three ground-state regimes in two dimensions as a result of increasing $\chi$ from zero: disordered, ``wavy crystalline," and crystalline (disordered and crystalline configurations were discovered in three dimensions, but wavy crystalline configurations were not, though they are postulated to exist).  Disordered ground states were found to minimize the potential energy for $\chi<0.5$ and are novel for classical systems (although disordered ground states for quantum mechanical systems have been shown to exist \cite{Torquato08_2}).  Wavy crystalline ground states appear for the range $0.5 < \chi < 0.7$ (although this was only demonstrated for two-dimensional systems).  Perfectly ordered crystalline states appear as the $\mathbf{k}$-space exclusion region comes close to the boundary of the second Brillouin zone, for $\chi\sim0.7$ in two dimensions.

It was found that if $\chi$ is large enough, the sole ground state for this class of potentials in two dimensions is the triangular lattice.  Initial theoretical studies proved this to be the case \cite{Suto05,Suto06,Torquato08,Theil06}.  Later numerical studies utilizing the collective-coordinate approach confirmed the theory.  Namely, it was shown that above a certain critical $\chi = \chi_c$, the unique ground state of a potential is the triangular lattice \cite{Uche06,Batten08}.  Below this critical value, however, the ground states remain degenerate.


\subsection{Sample directional potentials}

In the present paper, we  analyze the ground states for several
anisotropic or directional potentials with various $\mathbf{k}$-space exclusion regions.  For example, we studied the case where the exclusion region is an ellipse.  We explored the parameter space of the model varying both $\chi$, defined in the previous section, as well as $b/a$, the aspect ratio of the ellipse.  Various quantities of import were calculated, including the static structure factor, $S(\mathbf{k})$, which, as previously stated, describes the intensity of scattered  waves, as well as the pair correlation function, $g_2(\mathbf{r})$, which describes the conditional probability of finding a particle at position $\mathbf{r}$ given a particle fixed at the origin.  The numerical and analytical results for the ellipse potential are presented in the Appendix.

Two other potentials in two dimensions were analyzed, which will be mentioned here, briefly: the $\mathbf{k}$-space square and the $\mathbf{k}$-space plus sign.  Disordered ground state configurations for these potentials did not possess any noticeably unique properties.  Ordered ground state configurations were not found for either potential.  These potentials were not pursued any further, because their ground states did not display any qualitatively new features.

\subsection{Two-circle directional potential}

We focus our attention on the analysis of the ground states for directional
potentials whose $\mathbf{k}$-space exclusion region is two circles symmetric about the origin.  We define the dimensionless parameter, 
\begin{equation}
\alpha\equiv \frac{k_{disp}}{K},
\end{equation}
the ratio of the circles' displacement from the origin, $k_{disp}$, to their radius, $K$.  Once again, the parameter space was explored, varying both $\chi$ and $\alpha$.  The quantities $S(\mathbf{k})$ and $g_2(\mathbf{r})$ were also calculated.  See Figs. 1 and 2 for visualizations of this potential.

\begin{figure}[h!]
\centering
{\label{fig:1}\includegraphics[width=1\textwidth]{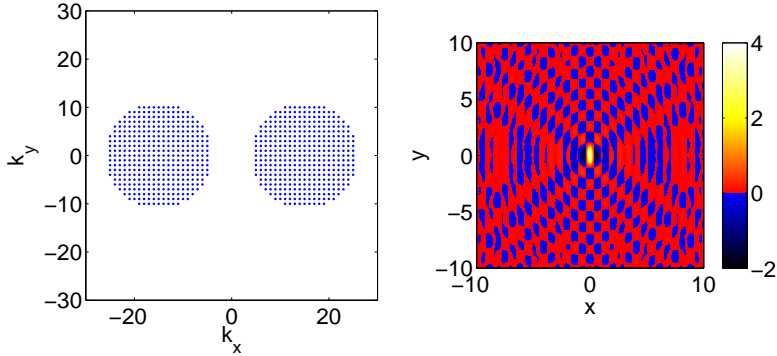}}
\caption{Two-circle potential, Fourier (left panel) and real-space (right panel) representations.  The dimensionless ratio of displacement distance to circle radius, $\alpha\equiv k_{disp}/K$, is $1.4548$.  The fraction of constrained degrees of freedom, $\chi$, is set to be $0.4461$. For these values, the particle soft core has an $x$-direction diameter of approximately 1/20 the box length and a $y$-direction diameter of approximately 1/10 the box length. The energy is scaled such that $V_o$ in Eq. (26) is unity (Color figure online)}
\end{figure}

\begin{figure}[h!]
\centering
{\label{fig:2}\includegraphics[width=1\textwidth]{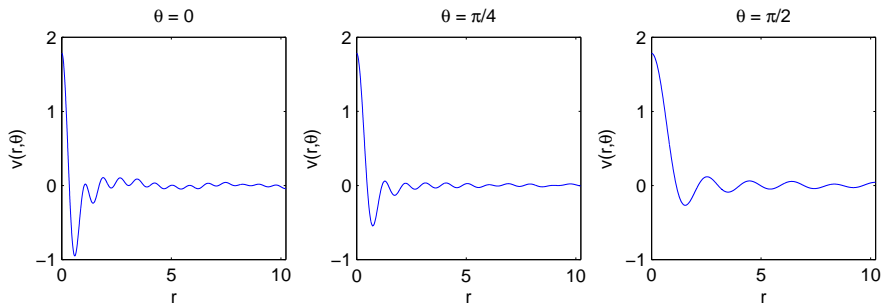}}
\caption{Real-space two-circle potential Eq. (26), $v(r,\theta)$ plotted radially (vs. $r$) at fixed angle, $\theta$.  The dimensionless ratio of displacement distance to circle radius, $\alpha\equiv k_{disp}/K$, is $1.4548$.  The fraction of constrained degrees of freedom, $\chi$, is set to $0.4461$.  The difference in the $x$ and $y$-direction soft core diameters can be observed. The energy is scaled as in Fig. 1.}
\end{figure}


\subsection{Simulation results}

Several ground state configurations were generated for all potentials, over a variety of parameters.  
As was the case in previous studies, convergence for $\chi<0.5$ was trivial for all potentials.  For the two-circle potential, on the other hand, convergence was obtained for all $\chi$ up to $\chi\sim0.74$ and for the entire tested range of the parameter $\alpha$.  Three different regimes were identified: disordered (with void regions), ``filamentary," and ``ordered" (with a collapse of dimensionality);  see Fig. 3 for examples of such configurations.  In general, it was observed that convergence was slower for the two-circle potential, likely having to do with the ``roughness" of the energy landscape, as discussed in the next section. Sample ground-state configurations for the two-circle potential, ranging from disordered to ordered, are presented below.

\begin{figure}[h!]
\centering
{\label{fig:3}\includegraphics[width=1.1\textwidth]{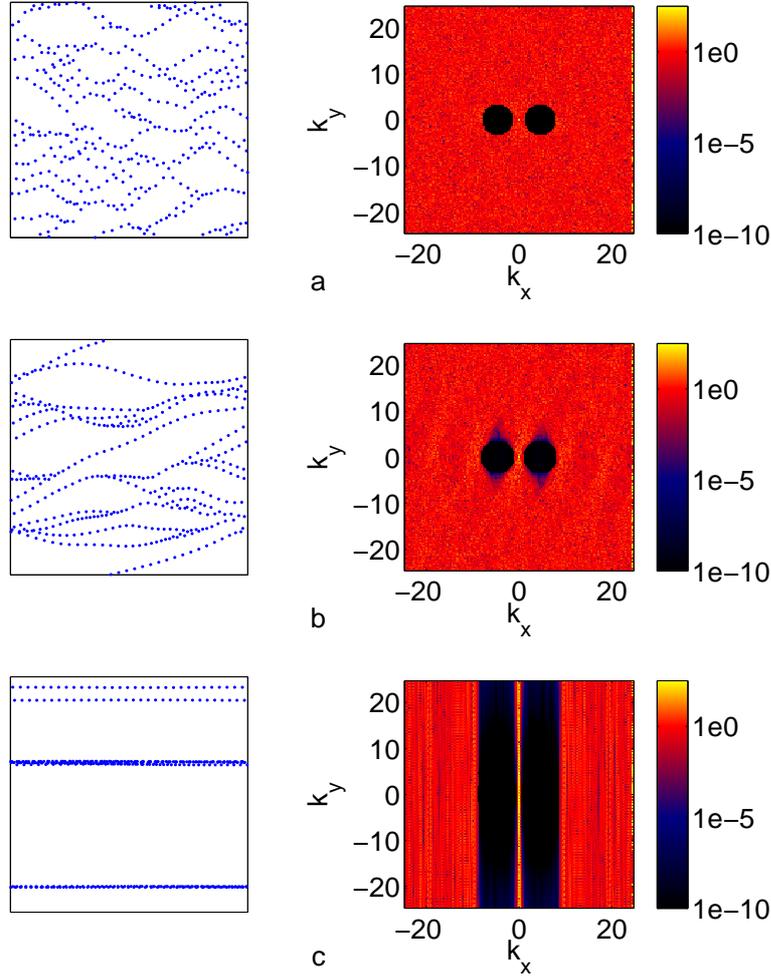}}
\caption{(a) Disordered, (b) ``filamentary," and (c) ordered configurations for the two-circle potential (Eq. 26), with fixed $k_{disp} = 4.8710$ (in units of inverse length) and $\alpha = 1.4548$ ($\chi=0.4461$), $\alpha = 1.3439$ ($\chi=0.5227$), and $\alpha = 1.1384$ ($\chi=0.7285$), respectively, with $N=418$ particles.  Within each row of the figure, the left image is a sample configuration and the right image is the logarithmically plotted structure factor $S(\mathbf{k})$.  Black regions indicate that $S(\mathbf{k})$ has been constrained to a minimum.  Lighter regions indicate that $S(\mathbf{k})>1$. (Color figure online)}
\end{figure}

\section{Two-circle potential: translation of constrained $C(\mathbf{k})$}

We go on to derive relations between the two-circle potential and the isotropic ``stealthy" potential in two-dimensions, which will aid in the analysis of our results.  In two-dimensional $\mathbf{k}$-space, we define the indicator function for some compact set $\mathcal{K}$ centered at the origin to be:
\begin{equation}
\Theta_\mathcal{K}(\mathbf{k}) = \Bigg\{\begin{matrix}
&1& \textrm{,   $\mathbf{k}\in \mathcal{K}$}\\
&0& \textrm{,   $\mathbf{k}\notin \mathcal{K}$}
 \end{matrix}\textrm{   }.
\end{equation}

\noindent We assume that this indicator function has a well-defined Fourier transform $f_K(\mathbf{r})$.  We proceed to translate the indicator function by some constant vector $\mathbf{k}_{disp}$ and its additive inverse $-\mathbf{k}_{disp}$, giving us $\Theta_K(\mathbf{k}-\mathbf{k}_{disp})$ and $\Theta_K(\mathbf{k}+\mathbf{k}_{disp})$, respectively.  We can signify the union of the translated sets by the sum of the indicator functions.  By the linearity of the Fourier transform operation, the Fourier transform of the union is then the sum of the Fourier transforms of the two translated indicator functions.  The shift property of the Fourier transform \cite{Stein03} gives us:
\begin{equation}
\mathcal{F}_\mathbf{r}[f(\mathbf{k}-\mathbf{k}_{disp})] = \mathcal{F}_\mathbf{r}[f(\mathbf{k})]e^{-2\pi i\mathbf{k}_{disp}\cdot\mathbf{r}}
\end{equation}

\noindent Putting this together, we see:
\begin{eqnarray}
\nonumber\mathcal{F}[\Theta_\mathcal{K}(\mathbf{k}-\mathbf{k}_{disp})+\Theta_\mathcal{K}(\mathbf{k}+\mathbf{k}_{disp})] & = & \mathcal{F}[\Theta_\mathcal{K}(\mathbf{k}-\mathbf{k}_{disp})]+\mathcal{F}[\Theta_\mathcal{K}(\mathbf{k}+\mathbf{k}_{disp})] \\
\nonumber& = & f_\mathcal{K}(\mathbf{r})e^{-2\pi i\mathbf{k}_{disp}\cdot\mathbf{r}} + f_\mathcal{K}(\mathbf{r})e^{2\pi i\mathbf{k}_{disp}\cdot\mathbf{r}}\\
& = & 2\textrm{cos}(2\pi\mathbf{k}_{disp}\cdot\mathbf{r})f_\mathcal{K}(\mathbf{r}).
\end{eqnarray}

Therefore, for the two-circle $\mathbf{k}$-space representation, the real-space potential will be that of the isotropic pair potential, modulated by a factor of $2\textrm{cos}(2\pi\mathbf{k}_{disp}\cdot\mathbf{r})$:
\begin{eqnarray}
v(\mathbf{r}) &=& \frac{V_oK}{2\pi r}\;\;J_1(Kr)\textrm{cos}(2\pi\mathbf{k}_{disp}\cdot\mathbf{r})\\
&=& \frac{V_oK}{2\pi r}\;\;J_1(Kr)\textrm{cos}(2\pi\alpha K\hat{\mathbf{k}}_{disp}\cdot\mathbf{r}),
\end{eqnarray}
where we have switched from radial to vectorial representation and where $\hat{\mathbf{k}}_{disp}$ is the unit vector in the direction of $\mathbf{k}_{disp}$.  The parameter $K = \sqrt{\chi/\pi}$ is the radius of the transformed circle.  The parameter $\alpha$ is as defined in Eq. (21) and the other parameters are as defined in Eq. (19).

We observe that the pair correlation function is equivalent to that of the ground states of the isotropic stealth potential for fixed angle $\theta=\pi/2$.  As the angle goes to $\theta=0$, we see the development of a large peak near $r=0$ and the emergence of long-ranged oscillations;
see Fig. 4 for an illustration of this behavior.  It was mentioned previously that the optimization procedure took significantly more time to find a ground state for all $\chi$ for the two-circle potential.  It can be hypothesized that this is the case because the added modulation makes the pair potential as well as the total energy landscape more ``rough" (i.e, many local minima),
thus. making the search for the absolute minimum more difficult.

\begin{figure}[h!]
\centering
{\label{fig:4}\includegraphics[width=1.1\textwidth]{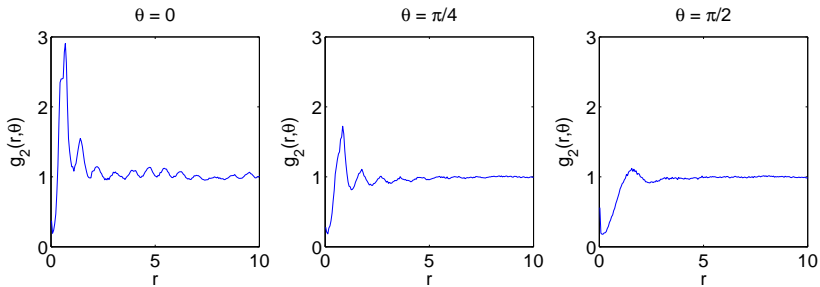}}
\caption{Plot of the pair correlation function $g_2(r,\theta)$ versus $r$ for selected angles ($\theta = 0,\,\pi/4,\,\pi/2$), $\chi = 0.4461$, $\alpha = 1.4548$, which has been averaged over 1000 configurations.  For $\theta=0$, the pair correlation function displays long-ranged, oscillatory behavior which disappears as $\theta\rightarrow\pi/2$.}
\end{figure}

\section{Discussion}

\subsection{Disordered ground states}
Ground-state configurations were generated at fixed $\chi$ values ($\chi<0.5$) while varying the parameter $\alpha$ for the two-circle potential.  For $\alpha>1$, it is apparent that the pair correlation function is modulated by a factor of cosine in the $x$-direction (see Figs. 6 and 7 for pair correlation and structure factor plots).  In addition, if $\chi$ is chosen from the approximate range $0.3<\chi<0.5$, noticeable void regions begin to emerge (calling to mind, though qualitatively, the voids present in the galaxy distribution \cite{Weygaert91}).  It was found that the shape and size of these void regions could be tuned by varying both parameters of the model.  For $\alpha\sim1$, the void regions display defined, polymer-like boundaries, which seem to be directed along the $x$-axis.  As $\alpha$ is increased, the void size stays approximately the same, while the distribution of particles around the voids seems to become more uniform (see Fig. 5).  This seems to be in accordance with the nature of the potential.  For smaller $\alpha$, the cosine modulation creates deep potential wells along the $x$-direction (refer to Fig. 2 for a visualization of this).  As $\alpha$ is increased, the oscillations of the cosine modulation become much more rapid along the axis, decreasing the width of the deepest wells, while introducing smaller wells with greater frequency in the $x$-direction.
It is useful to note that for $\alpha >1$, the two-circle potential is marginally stable (its volume 
integral or, equivalently. $V(k=0)$, is zero \cite{Ruelle}), which may help to explain the
existence of void regions. However,  this marginal stability property is a necessary but not
sufficient condition for the appearance of voids, since they only arise
for sufficiently large values of $\chi$ when $\alpha >1$.

\begin{figure}[h!]
\centering
{\label{fig:5}\includegraphics[width=1.1\textwidth]{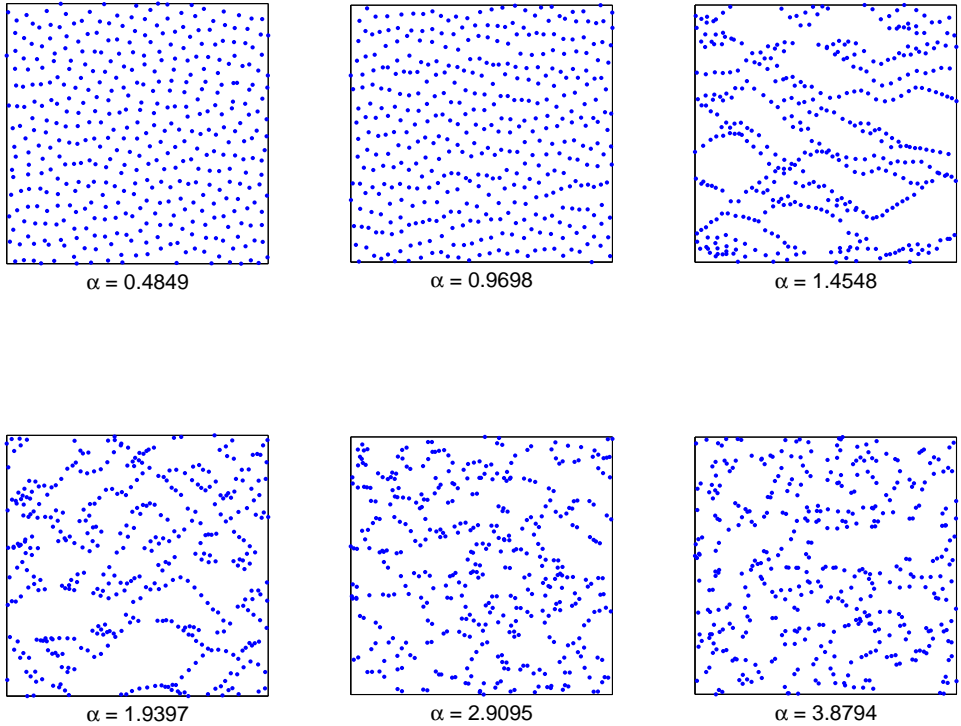}}
\caption{Disordered ground states for the two-circle potential defined by Eq. (26), varying the dimensionless parameter, $\alpha$, while keeping fixed $\chi = 0.4461$.  Large void regions emerge for the range $\alpha>1$.  The void properties noticeably change as this parameter is increased.}
\end{figure}

\begin{figure}[h!]
\centering
{\label{fig:6}\includegraphics[width=1.1\textwidth]{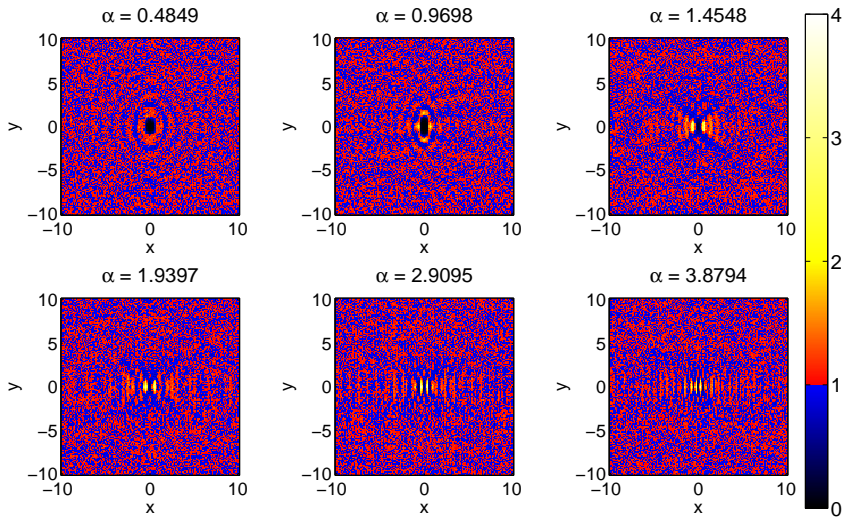}}
\caption{Two-dimensional pair correlation function for the two-circle potential Eq. (26) plotted in the $x$-$y$ plane as a function of dimensionless parameter $\alpha$, $\chi = 0.4461$.  The $\alpha$ values here correspond to the ones indicated in Fig. 5.  Black regions indicate anti-correlation, $g_2<1$.  Lighter regions indicate spatial correlation, $g_2>1$.  Each plot has been averaged over twenty configurations. (Color figure online)}
\end{figure}

\begin{figure}[h!]
\centering
{\label{fig:7}\includegraphics[width=1.1\textwidth]{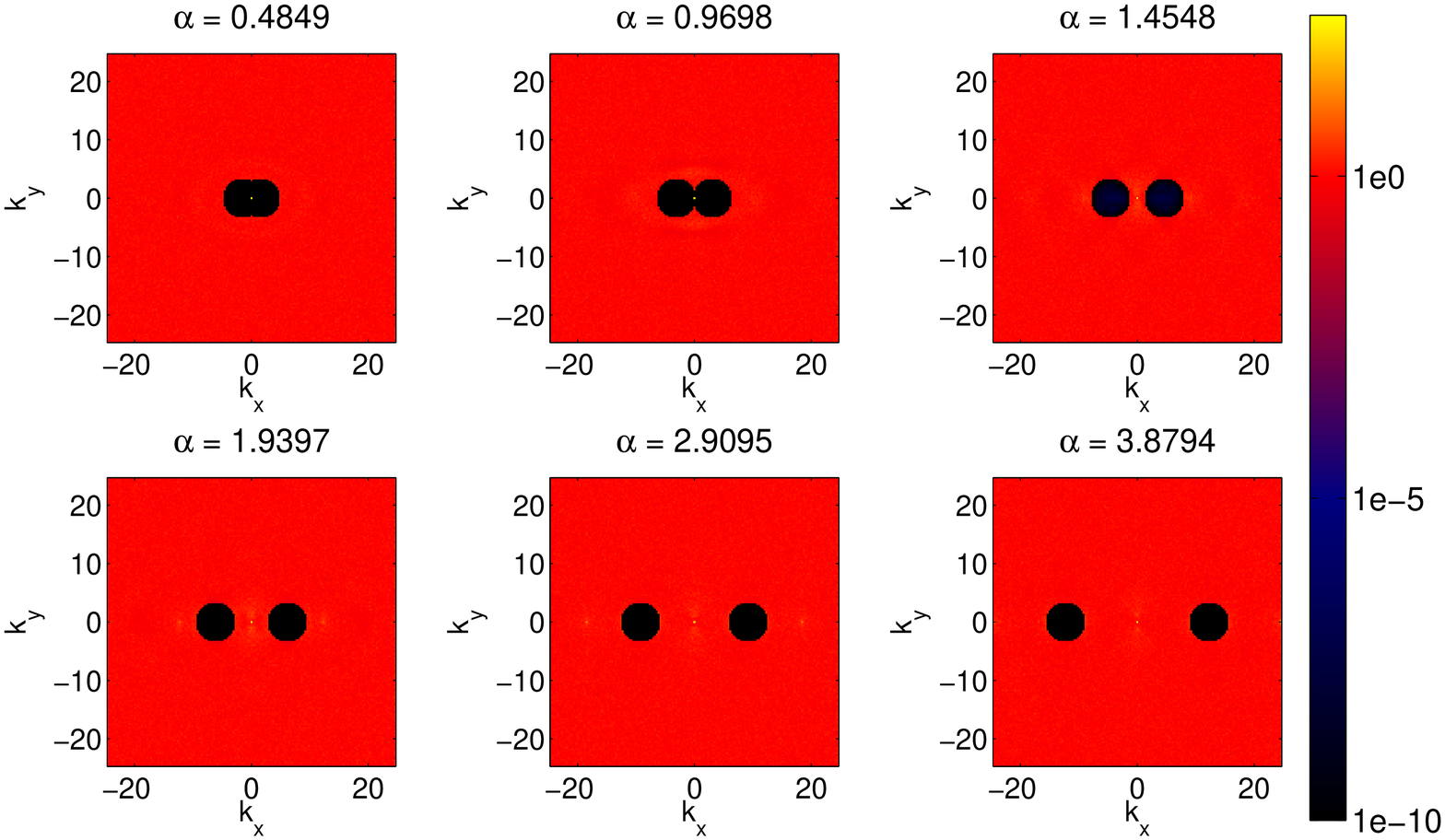}}
\caption{Two-dimensional static structure factor plotted logarithmically for the two-circle potential Eq. (26) as a function of dimensionless parameter $\alpha$ with fixed $\chi = 0.4461$.  The $\alpha$ values here correspond to the ones indicated in Fig. 5.  Black regions indicate that $S(\mathbf{k})$ has been constrained to zero (to within numerical error).  The lightest regions indicate that $S(\mathbf{k})$ has reached a maximum, $S(\mathbf{k})\sim N$, where $N$ is the number of particles.  Each plot has been averaged over twenty configurations. (Color figure online)}
\end{figure}


\subsection{Ordered ground states}
The ordered ground states of the two-circle potential system display surprising characteristics. For
$0.5<\chi<0.7$, ground-state configurations exhibit an increasing Òcollapse of dimensionalityÓ with
increasing $\chi$. For $\chi$ near 0.5, the particles fall into polymeric chains which are oriented in
the direction of displacement of the circles in the $\mathbf{k}$-space representation of the potential.
When $\chi$ is increased to its maximal value ($\chi\sim0.7$), the particles collapse to several nearly
perfectly linear chains, once again oriented in the direction of displacement. The chains are randomly
distributed along the $y$-axis and the number of linear chains increases as the parameter $\alpha$ is
increased. In some cases, the chains take on qualitatively sinusoidal shapes which are ordered along the
$x$-coordinate.

We now clarify exactly what we mean by ``collapse of dimensionality," referring to Fig. 3c.  Each one-dimensional linear chain (along the $x$-direction) is a minimal energy configuration of the potential in two dimensions.  Within each chain, however, the number of particles and relative order (in one dimension) vary, with the particles being subject to an effective one-dimensional potential along the $k_y$-direction with an effective $\chi$ value.  This effective potential arises as a result of the structure factor being minimized on a set of unconstrained wave vectors (i.e., the parts of the black regions in the structure factor plot of Fig. 3c which do not include the two circles).  Taking a slice along a fixed $k_y$ (for all $k_y$) gives the structure factor for a one-dimensional ground state subject to a potential which is a ``donut" in one-dimensional $k$-space, 
which is the aforementioned effective potential.  The two topmost chains in Fig. 3c, being ordered, have high effective $\chi$ values and the bottom two chains, being disordered, have low effective $\chi$ values in this reduced dimensional space.  The variation comes as a result of the fact that the union of any two ground-state configurations is itself a ground-state configuration and that these one-dimensional structures are similarly entropically favored.  Since the system is primarily constrained along the $x$-direction, the chains are free to be placed essentially randomly along the $y$-axis.

As the system collapses to these chain-like structures oriented along the $x$-direction, a pair of wedge-shaped regions along the collapsed direction in Fourier space (the $y$-direction) is forced to have positive $S(\mathbf{k})$; see Fig. 3c.  The angle of the wedge near the origin determines the nature of the filamentary structures.  The larger the wedge angle, the more tortuous are the filamentary structures.  As the wedge angle decreases, the filamentary structures are forced to become flat.  The presence of these wedges explains why these configurations do not arise for isotropic potentials such as the stealth potential \cite{Batten08}.

\subsection{Three classes of structures}
\subsubsection{Ordered chains}
For one such class of chains, those which display crystallinity in the direction of the chain, we are able to analyze the problem exactly.  We derive an expression for the structure factor of an ordered chain containing $m$ particles, parallel to the $x$-axis:
\begin{eqnarray}
\nonumber|\tilde{\rho}(\mathbf{k})|^2 &=& \bigg|\sum_{n=1}^me^{i\mathbf{k\cdot r}}\bigg|^2\\
\nonumber	&=& \bigg|e^{ik_yy_o}\sum_{n=1}^me^{ik_xx_on}\bigg|^2\\
\nonumber	&=& \bigg(\sum_{n=1}^me^{ik_xx_on}\bigg)\bigg(\sum_{n=1}^me^{-ik_xx_on}\bigg)\\
\nonumber	&=& \frac{\textrm{cos}(k_xx_om)-1}{\textrm{cos}(k_xx_o)-1}\\
			&=& \frac{\textrm{cos}(2\pi j_x)-1}{\textrm{cos}(2\pi j_x/m)-1}.
\end{eqnarray}
Therefore, there will be Bragg peaks at wave vectors $nm/(2\pi L)$, where $n$ is an integer, $m$ is the number of particles in the chain and $L$ is the length of the fundamental cell along the direction of the chain.

The form of the potential sets a bound on the number of particles which can form such a structure since no Bragg peaks can be found within the exclusion region.  We call the $k_x$-distance from the origin to the first Bragg peak $l$, the $k_x$-distance from the origin to the perimeter of the exclusion region $f$ and the diameter of the circle $d$.  There are then two regimes in which an ordered chain can exist:
\begin{eqnarray}
&l&>f+d\\
f>&l&>\frac{f+d}{2}.
\end{eqnarray}
The parameters $d$ and $f$ can be related to the parameters of the model, $\alpha$ and $\chi$, and $l$ can be related to the number of particles in the chain, giving us:
\begin{eqnarray}
&m&>\sqrt{\frac{\pi}{\chi}}(\alpha+2)\\
\alpha\sqrt{\frac{\pi}{\chi}}>&m&>\frac{1}{2}\sqrt{\frac{\pi}{\chi}}(\alpha+2).
\end{eqnarray}
A plot of an ordered chain and its structure factor is presented in Fig. 8, illustrating that the configuration minimizes the structure factor on wave vectors which fall within the two-circle exclusion region and is therefore a valid ground state.

\begin{figure}[h!]
\centering
{\label{fig:8}\includegraphics[width=1.1\textwidth]{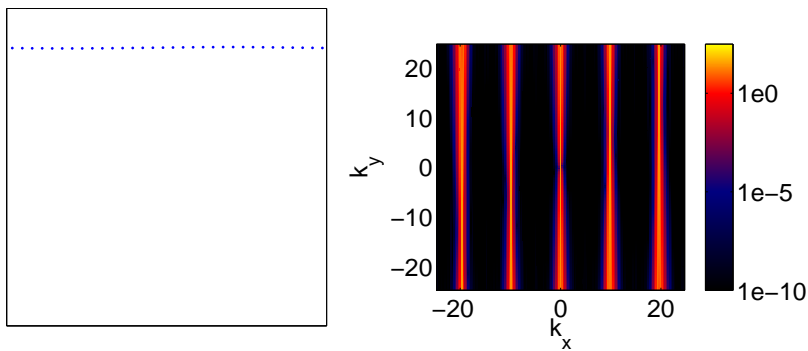}}
\caption{Sample ordered subconfiguration with $N=31$ particles (left panel) and its two-dimensional structure factor $S(\mathbf{k})$ plotted logarithmically.  Black regions indicate that $S(\mathbf{k})$ has been minimized. The structure minimizes $S(\mathbf{k})$ for the wave vectors within the non-zero region of the two-circle potential, indicating that the configuration is a ground state.  The thickness of the vertical lines present in the structure factor is notable, indicating that the chain is not perfectly aligned in the $y$-direction. (Color figure online)}
\end{figure}

\subsubsection{Ordered sinusoids}
We numerically calculate the structure factor for a representative sinusoid.  For a sinusoid to be a valid ground-state structure, its structure factor must be minimized on the set of wave vectors specified by the potential. A plot of the numerically calculated structure factor is presented in Fig. 9.  

\begin{figure}[h!]
\centering
{\label{fig:9}\includegraphics[width=1.1\textwidth]{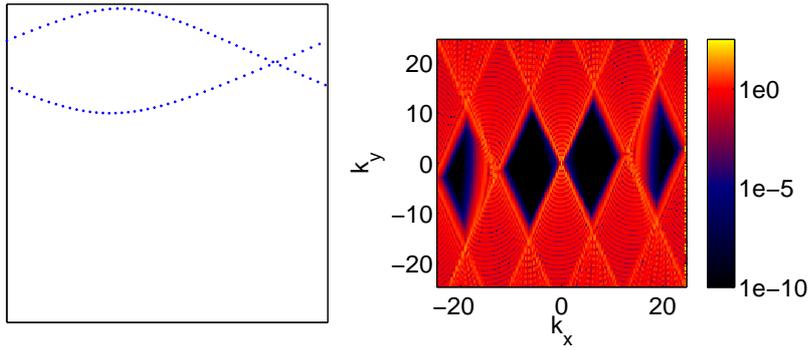}}
\caption{As in Fig. 8 except for a sample sinusoidal subconfiguration with $N=87$ particles.  Note that the period of the sinusoidal chain is twice the box length.  (Color figure online)}
\end{figure}

\subsubsection{Disordered chains}
Analytical expressions for the structure factor of the disordered chains were difficult to derive, so numerical methods were employed again.  A plot of the numerically calculated structure factor is presented in Fig. 10.  

\begin{figure}[h!]
\centering
{\label{fig:10}\includegraphics[width=1.1\textwidth]{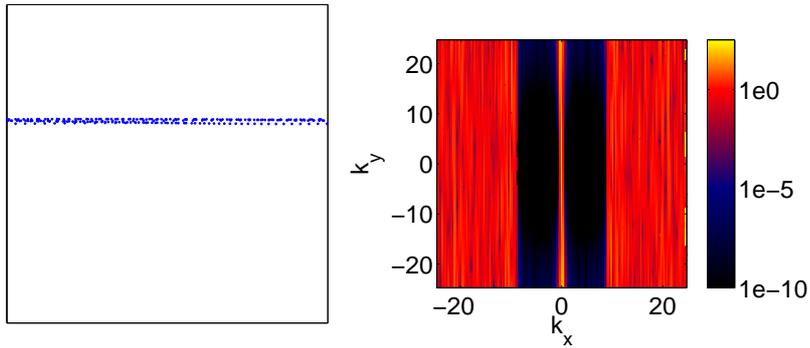}}
\caption{As in Figs. 8 and 9 except for a sample disordered configuration with $N=220$ particles.  (Color figure online)}
\end{figure}

\section{Concluding Remarks}
\label{conc}

The ground-state properties for classical many-particle systems subject to a directional pair interaction were investigated.  We found approximate relations for the pair statistics of the disordered ground states, which display large voids with a tunable structure.  The most ordered configurations display a collapse of dimensionality with the particles forming a few parallel, approximately linear, sometimes periodic chains in the $x$-direction with aperiodic placement of chains in the $y$-direction.  This behavior is new and completely unexpected given previous studies of the collective-coordinate potentials.  It should be noted, however, that so-called ``patchy particles," with directional real-space interactions
can also exhibit a collapse of dimensionality \cite{Giacometti10,Bianchi11}.  However, such particles possess internal rotational degrees of freedom that are absent in the many-particle systems that we consider in this paper.

We have explored the relations between isotropic and anisotropic pair potentials and their associated 
classical ground states, both here and in previous papers, highlighting the diversity produced by the 
collective-density variable approach \cite{Fan91,Uche04,Uche06,Batten08,Zachary11}.  However, many future avenues for research remain within these topics.  As previously noted, each interaction potential generates 
a highly degenerate classical ground state occupying a domain whose dimensionality depends on the 
number of collective coordinate constraints that have been imposed.  It will be important to devise 
mathematically precise descriptions of the connectivities, local curvatures, and configuration-space 
dispersions of these domains.  Furthermore, extension of the present study of the two-circle potential 
to a three-dimensional analog might shed light on the collapse of particle arrangement dimensionality 
in the plane described above.  Indeed, preliminary results for the corresponding ``two-sphere" potential 
model in three dimensions show once again that the particles exhibit a strong tendency to organize 
themselves into directed filamentary structures.

\appendix

\section{Affine Theorem and the Ellipse Potential}
\label{Affine}
We derive a simple relation transform describing how the Fourier transform of a function in $d$-dimensional Euclidean space $\mathbb{R}^d$ behaves under invertible affine transformations. We also present selected  results for the ellipse potential described in Sec. III.  The ellipse potential is described by the dimensionless aspect ratio, $b/a$, where $b$ is the length of the minor axis and $a$ is the length of the major axis of the potential in $\mathbf{k}$-space.  We go on to show that the ellipse potential is related to the previously studied circle potential by a simple affine transformation, which allows for a simple analysis of the ground state patterns.

\subsection{Affine Theorem for the Fourier Transform}
We derive a simple relation transform describing how the Fourier transform behaves under invertible affine transformations, generalizing a result in two-dimensions to arbitrary dimension \cite{Bracewell93}.  Given a square-integrable function $f(\mathbf{r})$ in $d$-dimensional Euclidean space, we can define its Fourier transform as:
\begin{equation}
F(\mathbf{k}) = \mathcal{F}_\mathbf{k}[f(\mathbf{r})] = \int f(\mathbf{r})e^{-i\mathbf{k}\cdot\mathbf{r}}d\mathbf{r}.
\end{equation}
The inverse Fourier transform is then naturally defined to be:
\begin{equation}
f(\mathbf{r}) = \mathcal{F}^{-1}_\mathbf{r}[F(\mathbf{k})] = \frac{1}{(2\pi)^d}\int F(\mathbf{k})e^{i\mathbf{k}\cdot\mathbf{r}}d\mathbf{k}.
\end{equation}
Let $f(\mathbf{r})$ be a function, $f: \mathbb{R}^d\rightarrow\mathbb{R}$ with a known Fourier transform $\mathcal{F}[f(\mathbf{r})] = \hat{f}(\mathbf{k})$. Let us apply an invertible linear transformation, $A$, to $\mathbf{r}\in\mathbb{R}^d$, and denote the inverse by $A^{-1}$.  In order to be completely general, we also apply a translation denoted by the addition of a vector $\mathbf{b}$.  We now have:
\begin{align*}
\mathcal{F}[f(A\mathbf{r}+\mathbf{b})] = \int d\mathbf{r}\;\;e^{-i\mathbf{k^Tr}}f(A\mathbf{r}+\mathbf{b}),
\end{align*}
where we have switched to linear algebra notation for ease and clarity of presentation.  Let $\mathbf{r}'\equiv A\mathbf{r}$.  In transforming the equation into the new coordinate system, we gain a factor $\frac{1}{\textrm{det}(A)}$ from the Jacobian.
\begin{align*}
\mathcal{F}[f(A\mathbf{r}+\mathbf{b})] = \frac{1}{\textrm{det}(A)}\int d\mathbf{r}'\;\;e^{i\mathbf{k^T}A^{-1}\mathbf{r}'}f(\mathbf{r}'+\mathbf{b}).
\end{align*}
Let $\mathbf{k}'\equiv (\mathbf{k^T}A^{-1})^T = (A^{-1})^T\mathbf{k}$.
\begin{eqnarray}
\nonumber\mathcal{F}[f(A\mathbf{r}+\mathbf{b})] 	&=& \frac{1}{\textrm{det}(A)}\int d\mathbf{r}'\;\;e^{i\mathbf{k'^Tr'}}f(\mathbf{r}'+\mathbf{b})\\
\nonumber						&=& \frac{1}{\textrm{det}(A)}e^{i\mathbf{k'^T}\mathbf{b}}\int d\mathbf{r}'\;\;e^{i\mathbf{k'^Tr'}}f(\mathbf{r}')\\
\nonumber					 	&=& \frac{1}{\textrm{det}(A)}e^{i\mathbf{k'^T}\mathbf{b}}\hat{f}(\mathbf{k}')\\
					 			&=& \frac{1}{\textrm{det}(A)}e^{i\mathbf{b^T}(A^{-1})^\mathbf{T}\mathbf{k}}\hat{f}((A^{-1})^T\mathbf{k}),
\end{eqnarray}
where the second step follows from the Fourier shift property \cite{Stein03}.  To the best of our knowledge, this result is the first of its kind.  It is consistent with the previously mentioned two-dimensional result, as well as the standard results corresponding to multiplication by a constant factor and rotations \cite{Stein03}.

Since our simulations are carried out in two dimensions, we will explicate the details of the previous result in this setting.  For some invertible affine transformation in two dimensions:
\begin{align*}
\mathbf{r}\rightarrow A\mathbf{r}+\mathbf{r}_\textrm{o},
\end{align*}
where $A$ is a linear transformation given by the invertible matrix:
\begin{align*}
A = 
\begin{bmatrix}
a & b \\
c & d
\end{bmatrix}
\end{align*}
and $\mathbf{r}_\textrm{o}$ is a translation vector:
\begin{align*}
\mathbf{r}_\textrm{o} = 
\begin{bmatrix}
f \\
g
\end{bmatrix},
\end{align*},
\noindent the Fourier transform of $f(A\mathbf{r})$ is:
\begin{equation} \label{eq_a}
\frac{1}{\Delta}e^{i\mathbf{k}\cdot\mathbf{r}_\textrm{o}}\hat{f}(A'\mathbf{k}),
\end{equation}

\noindent where $\hat{f}(\mathbf{k})$ is the Fourier transform of $f(\mathbf{r})$, $A'$ is a linear transformation defined by the matrix:
\begin{equation} \label{eq_b}
A' = \frac{1}{\Delta}\begin{bmatrix} 
d & -c \\
-b & a
\end{bmatrix}
\end{equation}
\noindent and $\Delta = \mathrm{det}(A) = ad-bc$.

In the case of the ellipse potential, the region on which $C(\mathbf{k})$ is constrained can be expressed as a circular exclusion region to which we have applied invertible affine transformations.  Since the ground-state properties of the circular exclusion region are well known (in light of previous studies), we can use this information, combined with this affine relation to draw conclusions about the ground-state properties of the ellipse potential.  We emphasize that the affine transformation is not applied to the particles themselves and therefore does not allow for the particles to rotate.

\begin{figure}[h!]
\centering
{\label{fig:11}\includegraphics[width=0.5\textwidth]{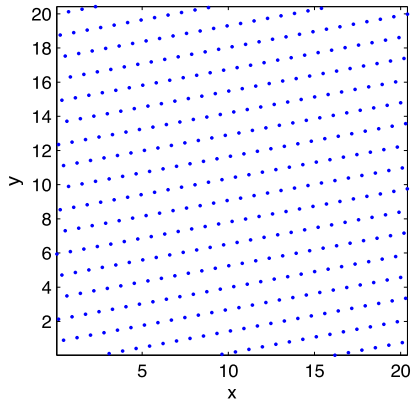}}
\caption{Crystalline ground state configuration for the ellipse potential.  If the configuration is subject to the squeeze transformation described above, the triangular lattice is not recovered.}
\end{figure}

\subsection{Results}
\subsubsection{Disordered ground states}
Ground state configurations were generated over a range of aspect ratios for the ellipse potential at a fixed $\chi$ value.  Comparing the pair correlation function of the ground states of a potential of given aspect ratio to that of the isotropic circle potential, it is apparent that the approximate affine relation derived in the previous section holds.  In addition, if $\chi$ is increased to values near but still below $0.5$, the particles develop what resemble hard cores and we see the emergence of perfect nematic order (which is by construction).
\subsubsection{Crystalline ground states}
One might hypothesize that the crystalline ground states would obey the same affine relationship that the potential exactly obeys and the pair statistics approximately obey, i.e., that the absolute ground state would be an appropriately transformed triangular lattice.  However, this is not the case.  As can be visually ascertained from Fig. 11, applying the inverse squeeze transformation does not recover the triangular lattice.  In order to confirm this, a perturbed transformed lattice configuration was used as an initial particle configuration for simulations using the ellipse potential.  Not only did the trials not find the desired lattice to be the ground state, but no true ground state was reached when these initial conditions were used.  However, it should be noted that 418 particles cannot be used to created a nearly undeformed, affine transformed triangular lattice in a square box (as is possible with the untransformed triangular lattice), which may have influenced the convergence of the procedure.

\begin{acknowledgements}
We are pleased to offer this contribution to honor M. Fisher, J. Percus, and B. Widom whose own remarkable contributions have vividly demonstrated the originality intrinsic to statistical mechanics.  This work was supported by the Office of Basic Energy Sciences, U.S. Department of Energy, under Grant No. DE-FG02-04-ER46108.
\end{acknowledgements}


\begin{thebibliography}{}
%
%
\bibitem{Feynman72}
Feynman, R.P.: Statistical Mechanics.
\newblock Benjamin, Reading, MA (1972).

\bibitem{Yevick56}
Yevick, G.~J., Percus, J.~K.: New approach to the many-body problem.
\newblock {\em Phys. Rev.} {\bf101}(3), 1186 (1956).

\bibitem{Percus58}
Percus, J.~K., Yevick, G.~J.: Analysis of Classical Statistical Mechanics by Means of Collective Coordinates.
\newblock {\em Phys. Rev.} {\bf110}:1 (1958).

\bibitem{Fan91}
Fan, Y., Percus, J.~K., Stillinger, D.~K., Stillinger, F.~H.: Constraints on collective density variables: One dimension.
\newblock {\em Phys. Rev. A} {\bf44}(4), 2394 (1991).

\bibitem{Uche04}
Uche, O.U., Stillinger, F.H., Torquato, S.: Constraints on collective density variables: Two dimensions.
\newblock {\em Phys. Rev. E} {\bf70}, 046122 (2004).

\bibitem{Suto05}
S\"ut\H{o}, A.: Crystalline Ground States for Classical Particles.
\newblock {\em Phys. Rev. Letters} {\bf95}, 265501 (2005).

\bibitem{Suto06}
S\"ut\H{o}, A.: From bcc to fcc: Interplay between oscillating long-range and repulsive short-range forces.
\newblock {\em Phys. Rev. B} {\bf74}, 104117 (2006).

\bibitem{Uche06}
Uche, O.U., Torquato, S. Stillinger, F.H.: Collective coordinate control of density distributions.
\newblock {\em Phys. Rev. E} {\bf74}, 031104 (2006).

\bibitem{Torquato08}
Torquato, S., Stillinger, F.H.: New Duality Relations for Classical Ground States.
\newblock {\em Phys. Rev. Letters} {\bf100}, 020602 (2008).

\bibitem{Batten08}
Batten, R.D., Stillinger, F.H., Torquato, S.: Classical disordered ground states: Super-ideal gases and stealth and equi-luminous materials.
\newblock {\em J. Appl. Phys.} {\bf104}, 033504 (2008).


\bibitem{Zachary11}
Zachary, C.E., Torquato, S.: Anomalous local coordination, density fluctuations, and void statistics in disordered hyperuniform many-particle ground states.
\newblock {\em Phy. Rev. E} {\bf83}, 051133 (2011).


\bibitem{Torquato11}
Torquato, S, Zachary, C.E., Stillinger, F.H.: Duality relations for the classical ground states of soft-matter systems.
\newblock {\em Soft Matter} {\bf 7}, 3780 (2011).

\bibitem{Batten11}
Batten, R. D., Stillinger, F.H., Torquato, S.: Inherent structures for soft long-range interactions in two-dimensional many-particle systems.
\newblock {\em J. Chem. Phys.} {\bf135}, 054104 (2011).

\bibitem{Likos06}
Likos, C.N.: Going to ground.
\newblock {\em Nature} {\bf 440}, 7083 (2006).

\bibitem{Zachary08}
Zachary, C.E., Stillinger, F.H., Torquato, S.: Gaussian core model phase diagram and pair correlations
in high Euclidean dimensions.
\newblock {\em J. Chem. Phys.} {\bf 128}, 224505 (2008).

\bibitem{Flory50}
Flory, P.J., Krigbaum, W.R.: Statistical mechanics of dilute polymer solutions. II.
\newblock {\em J. Chem. Phys.} {\bf 18}, 1086 (1950).

\bibitem{Venables00}
J. Venables,
\newblock {\em Introduction to Surface and Thin Film Processes}.
\newblock Cambridge UP, Cambridge, UK (2000).

\bibitem{Florescu09}
Florescu, M., Torquato, S., Steinhardt, P.J.: Designer disordered materials with large, complete photonic band gaps.
\newblock {\em PNAS} {\bf106}, 49 (2009).


\bibitem{Giacometti10}
Giacometti, A., Lado, F., Largo, J., Pastore, G., Sciortino, F.: Effects of patch size and number within a simple model of patchy colloids. {\em J. Chem. Phys.} {\bf132}, 174110 (2010).

\bibitem{Bianchi11}
Bianchi, E., Blaak, R., Likos, C.: Patchy colloids: state of the art and perspectives. {\em PCCP} {\bf 13} (2011).


\bibitem{Kaufman99}
Kaufman, L.: A reduced storage quasi-Newton trust region approach to function optimization
\newblock {\em SIAM J. Optim.} {\bf10}, 56 (1999).

\bibitem{To03}
Torquato, S.,  Stillinger, F. H: Local Density Fluctuations, Hyperuniform Systems, and Order Metrics, 
{\em Phys. Rev. E} {\bf 68}, 041113  (2003).

\bibitem{Torquato08_2}
Torquato, S., Scardicchio, A., Zachary, C.E.: Point processes in arbitrary dimension from fermionic gases, random matrix theory, and number theory.
\newblock {\em J. Stat. Mech. Theor. Exp.} P11019 (2008).

\bibitem{Theil06}
Theil, F.: A proof of crystallization in two dimensions.
\newblock {\em Comm. Math. Phys.} {\bf262}, 209 (2006).

\bibitem{Stein03}
Stein, E.M., Shakarchi, R.:
\newblock {\em Fourier Analysis: An Introduction},
\newblock Princeton UP, Princeton, (2003).

\bibitem{Weygaert91}
van~de Weygaert, R.:
\newblock {\em Voids and the Geometry of Large Scale Structure},
\newblock University of Leiden, Leiden, (1991).

\bibitem{Ruelle}
Ruelle, D., {\it Statistical Mechanics: Rigorous Results}, World Scientific, London, 1999.

\bibitem{Bracewell93}
Bracewell, R.N., Chang, K.-Y., Jha, A.K., Wang, Y.-H.: Affine theorem for two-dimensional Fourier transform.
\newblock {\em Electronics Letters} {\bf29}, 3 (1993).
%

\end{thebibliography}



\end{document}